\documentclass[aps,preprintnumbers,superscriptaddress,showpacs]{revtex4}
\usepackage{epsfig}
\usepackage{psfrag}
\usepackage{amsfonts}
\usepackage{graphicx}
\usepackage{dcolumn}
\usepackage{bm}

\begin{document}

%%%%% title page %%%%%
\title{Heavy quark potential and jet quenching parameter in a D-instanton background}

\author{Zi-qiang Zhang}
\email{zhangzq@cug.edu.cn} \affiliation{School of mathematics and
physics, China University of Geosciences(Wuhan), Wuhan 430074,
China}

\author{De-fu Hou}
\email{houdf@mail.ccnu.edu.cn} \affiliation{Institute of Particle
Physics and Key Laboratory of Quark and Lepton Physics (MOS),
Central China Normal University, Wuhan 430079,China}

\author{Gang Chen}
\email{chengang1@cug.edu.cn} \affiliation{School of mathematics
and physics, China University of Geosciences(Wuhan), Wuhan 430074,
China}

%%%%%%%%%%%%%%%%%%%%%%%%%%%%%%%%%%%%%%%%
\begin{abstract}
Using the AdS/CFT correspondence, we study the heavy quark
potential and the jet quenching parameter in the near horizon
limit of D3-D(-1) background. The results are compared with those
of conformal cases. It is shown that the presence of instantons
tends to suppress the heavy quark potential and enhance the jet
quenching parameter.
\end{abstract}

\pacs{12.38.Lg, 12.38.Mh, 11.25.Tq}

\maketitle
%%%%%%%%%%%%%%%%%%%%%%%%%%%%%%%%%%%%%%%%
\section{Introduction}

One main purpose of the heavy-ion collision experiments is to
explore the properties of the new state of matter created through
collisions. The experiments at RHIC and LHC have produced a new
state of matter so-called "strong quark-gluon plasma(sQGP)"
\cite{JA,KA,EV}. Thus, non-peturbative techniques are required
such as the AdS/CFT correspondence
\cite{Maldacena:1997re,Gubser:1998bc,MadalcenaReview}.

AdS/CFT, the duality between the type IIB superstring theory
formulated on AdS$_5\times S^5$ and $\mathcal N=4$ SYM in four
dimensions, has yielded many important insights into the dynamics
of strongly-coupled gauge theories. In this approach, many
quantities such as the heavy quark potential and the jet quenching
parameter can be studied.

The heavy quark potential is an important quantity which can be
related to the melting of heavy quarkoniums, one of the main
experimental signatures for sQGP formation. The first calculation
of the heavy quark potential for $\mathcal N=4$ SYM at zero
temperature was carried out by Maldacena \cite{Maldacena:1998im}.
It was observed that for the $AdS_5$ space the energy shows a
purely Coulombian behavior, agrees with a conformal gauge theory.
This work has attracted lots of interest. After
\cite{Maldacena:1998im}, the heavy quark potential in the context
of AdS/CFT has been investigated in many papers. For example, the
potential for $\mathcal N=4$ SYM at finite temperature has been
discussed in \cite{AB,SJ}. The potential for different spaces is
investigated in \cite{YK}. The sub-leading order corrections to
this quantity are considered in \cite{SX} and \cite{ZQ}. For study
of the potential in some AdS/QCD models, see \cite{OA2,SH,SH1}.
Other important results can be found, for example, in
\cite{JG,FB,LM,KB1}.

Another important quantity sensitive to the in-medium energy loss
is jet quenching parameter $\hat{q}$(or transport coefficient).
This quantity describes the average transverse momentum square
transferred from the traversing parton, per unit mean free path
\cite{RB,XN}. The jet quenching parameter for $N=4$ SYM theory was
first proposed by H.Liu et al in their seminal work \cite{liu}.
Interestingly, the magnitude of $\hat q_{SYM}$ turns out to be
closer to the value extracted from RHIC data \cite{K.J,A.D} than
pQCD result for the typical value of the 't Hooft coupling,
$\lambda\simeq 6\pi$, of QCD. After \cite{liu}, there are many
attempts to address the jet quenching parameter from AdS/CFT. For
instance, the sub-leading order corrections to $\hat{q}$ due to
worldsheet fluctuations has been discussed in \cite{ZQ1}. Charge
effect and finite 't Hooft coupling correction on the $\hat{q}$ is
investigated in \cite{KB}. The $\hat{q}$ in medium with chemical
potential is studied in \cite{FL,NA,SD}. The jet quenching
parameter in STU background is analyzed in \cite{KB2}.
Investigations are also extended to some AdS/QCD models
\cite{EN,UG}. Other related results can be found, for example, in
\cite{AF1,AB1,JF,EC,LH,MB,ZQ2}.

Actually, there is another check of gauge/gravity duality, the
correspondence between non-perturbative objects such as
instantons. It was argued \cite{ND,OA} that the Yang-Mills
instantons are identified with the D-instantons of type IIB string
theory. The near horizon limit of D-instantons homogeneously
distributed over D3-brane at zero temperature has been discussed
in \cite{LHH}. The holographic dual of uniformly distributed
D-instantons over D3-brane at finite temperature has been
investigated in \cite{BG}. It is shown that the features of
D3-D(-1) configuration is similar to QCD at finite temperature.
Therefore, one can expect the results obtained from this theory
should shed qualitative insights into analogous questions in QCD.
In this paper, we are going to study the heavy quark potential and
the jet quenching parameter in a D-instanton background. We will
investigate the effect of the instanton density on these two
quantities. Moreover, we would like to compare the results with
those of conformal cases and experimental data. This is the
purpose of the present work.

This paper is organized as follows. In the next section, the
background geometry of D3-D(-1) brane configuration at finite
temperature is briefly reviewed. In section III, we investigate
the heavy quark potential in this background. Then we study the
jet quenching parameter in this background in section IV. The last
part concludes the paper along with some discussions of the
results.

\section{D-Instanton Background}
Let us begin with a brief review of the D-instanton background.
The geometry is a finite temperature extension of D3/D-instanton
background with Euclidean signature \cite{KG}. The background has
a five-form field strength and a axion field which couples to D3
and D-instanton, respectively. The ten dimensional super-gravity
action in Einstein frame is \cite{GW,AK}
\begin{equation}
S=\frac{1}{\kappa}\int
d^{10}x\sqrt{g}(R-\frac{1}{2}(\partial\Phi)^2+\frac{1}{2}e^{2\Phi}(\partial\chi)^2-\frac{1}{6}F^2_{(5)}),\label{action}
\end{equation}
where $\Phi$ is the dilaton, $\chi$ denotes the axion. By setting
$\chi=-e^{-\Phi}+\chi_0$, the dilaton term and the axion term can
cancel. Then the solution with metric in string frame can be
written as \cite{BG}
\begin{equation}
ds^2_{10}=e^{\frac{\Phi}{2}}[-\frac{r^2}{R^2}f(r)dt^2+\frac{r^2}{R^2}d\vec{x}^2+\frac{1}{f(r)}\frac{R^2}{r^2}dr^2+R^3d\Omega^2_5],\label{metric}
\end{equation}
with
\begin{equation}
e^\Phi=1+\frac{q}{r^4_t}log\frac{1}{f(r)},\qquad
\chi=-e^{-\Phi}+\chi_0,\qquad f(r)=1-\frac{r_t^4}{r^4},
\end{equation}
where R is the radius of curvature, $\vec{x}$ stands for the
spatial directions of the space time, r denotes the radial
coordinate of the geometry, $r_t$ is the radius of the event
horizon. The parameter $q$ refers to the number of D-instanton. In
the framework of AdS/CFT duality, $q$ also represents the vacuum
expectation value of gluon condensation \cite{BG}.

The Hawking temperature of the black hole is given by
\begin{equation}
T=\frac{r_t}{\pi R^2}.
\end{equation}

\section{heavy quark potential}

In this section, we study the heavy quark potential in a
D-instanton background. The heavy quark potential can be extracted
from the expectation value of the following Wilson loop
\begin{equation}
W(C)=\frac{1}{N_c}Tr Pe^{ig\oint_C dx^\mu A_\mu},
\end{equation}
where C refers to a closed loop in space time and the trace is
over the fundamental representation of SU(N) group. $A_\mu$ is the
gauge potential. $P$ enforces the path ordering along the loop
$C$. The rectangular loop is along the time $\mathcal {T}$ and
spatial extension L.

The heavy quark potential is related to the expectation value of
W(C) in the limit $\mathcal {T}\rightarrow\infty$,
\begin{equation}
<W(C)>\sim e^{-\mathcal {T}V(L)}.
\end{equation}

On the other hand, according to the AdS/CFT duality, the
expectation value of W(C) is given by
\begin{equation}
<W(C)>\sim e^{-S_c},
\end{equation}
where $S_c$ is the regularized action which can be derived from
the Nambu-Goto action.

As a result, the heavy quark potential is expressed as
\begin{equation}
V(L)=\frac{S_c}{\mathcal {T}}\label{v}.
\end{equation}

We now to consider the heavy quark potential in the D-instanton
background by using the metric (\ref{metric}). The string action
can reduce to the Nambu-Goto action
\begin{equation}
S=-\frac{1}{2\pi\alpha^\prime}\int d\tau d\sigma\sqrt{-det
g_{\alpha\beta}},
\end{equation}
with
\begin{equation}
g_{\alpha\beta}=G_{\mu\nu}\frac{\partial
X^\mu}{\partial\sigma^\alpha} \frac{\partial
X^\nu}{\partial\sigma^\beta},
\end{equation}
where $\frac{1}{2\pi\alpha^\prime}$ denotes the string tension,
$\alpha^\prime$ is related to the 't Hooft coupling constant
$\lambda$ by
\begin{equation}
\frac{R^2}{\alpha^\prime}=\sqrt{\lambda},
\end{equation}
and $G_{\mu\nu}$ and $X^\mu$ are the metric and the target space
coordinates, $\sigma^\alpha$ parameterize the world sheet with
$\alpha=0,1$.

By using the static gauge
\begin{equation}
t=\tau, \qquad x^1=\sigma,
\end{equation}
and supposing that the radial direction only depends on $\sigma$,
\begin{equation}
r=r(\sigma),
\end{equation}
then the Euclidean version of Nambu-Goto action in (\ref{metric})
can be written as
\begin{equation}
S=\frac{\mathcal {T}}{2\pi\alpha^\prime}\int
d\sigma\sqrt{e^{\Phi}[f(r)\frac{r^4}{R^4}+\dot{r}^2]}.\label{ac}
\end{equation}

We now identify the Lagrangian as
\begin{equation}
\mathcal
L=\sqrt{e^{\Phi}[f(r)\frac{r^4}{R^4}+\dot{r}^2]},\label{L}
\end{equation}
where $\dot{r}=\frac{dr}{d\sigma}$.

Note that $\mathcal L$ does not depend on $\sigma$ explicitly, so
we have a conserved quantity,
\begin{equation}
\frac{\partial\mathcal L}{\partial\dot{r}}\dot{r}-\mathcal
L=constant.
\end{equation}

The boundary condition at $\sigma=0$ is:
\begin{equation}
\dot{r}=0, \qquad r=r_c, \qquad (r_t<r_c),
\end{equation}
which yields
\begin{equation}
\frac{e^\Phi
\frac{r^4}{R^4}f(r)}{\sqrt{e^\Phi[f(r)\frac{r^4}{R^4}+\dot{r}^2]}}=const=\sqrt{e^{\Phi
(r_c)}f(r_c)\frac{r_c^4}{R^4}},
\end{equation}
with
\begin{equation}
f(r_c)=1-\frac{r_t^4}{r_c^4},\qquad
e^{\Phi(r_c)}=1+\frac{q}{r_t^4}log\frac{1}{f(r_c)},
\end{equation}
then a differential equation is derived,
\begin{equation}
\dot{r}=\frac{dr}{d\sigma}=\frac{1}{R^2}\sqrt{\frac{(r^4-r^4_t)[e^\Phi(r^4-r_t^4)-e^{\Phi(r_c)}(r_c^4-r_t^4)]}{e^{\Phi(r_c)}(r_c^4-r_t^4)}}\label{dotr}.
\end{equation}

By integrating (\ref{dotr}), the distance between the
quark-antiquark pair is obtained
\begin{eqnarray}
L=2R^2\int_{r_c}^{\infty}dr\sqrt{\frac{e^{\Phi(r_c)}(r_c^4-r_t^4)}{(r^4-r^4_t)[e^\Phi(r^4-r_t^4)-e^{\Phi(r_c)}(r_c^4-r_t^4)]}}
.\label{x}
\end{eqnarray}

Substituting (\ref{dotr}) into (\ref{ac}), one finds the action of
the quark pair
\begin{equation}
S=\frac{\mathcal {T}}{\pi\alpha'}\int_{r_c}^{\infty}dr
\sqrt{\frac{e^{2\Phi}(r^4-r_t^4)}{e^\Phi(r^4-r_t^4)-e^{\Phi(r_c)}(r_c^4-r_t^4)}}.
\label{NG1}
\end{equation}

This action needs to be subtracted by $S_0$
\begin{equation}
S_0=\frac{\mathcal
{T}}{\pi\alpha'}\int_{r_t}^{\infty}dr\sqrt{e^\Phi}. \label{NG2}
\end{equation}

Thus, the regularized action is given by
\begin{equation}
S_c=S-S_0 \label{NG2}.
\end{equation}

Applying (\ref{v}), we end up with the heavy quark potential in a
D-instanton background
\begin{eqnarray}
V(L)=\frac{1}{\pi\alpha'}\int_{r_c}^{\infty}dr[
\sqrt{\frac{e^{2\Phi}(r^4-r_t^4)}{e^\Phi(r^4-r_t^4)-e^{\Phi(r_c)}(r_c^4-r_t^4)}}-\sqrt{e^\Phi}]-\frac{1}{\pi\alpha'}\int_{r_t}^{r_c}dr\sqrt{e^\Phi}\label{V0}.
\end{eqnarray}

\begin{figure}
\centering
\includegraphics[width=8cm]{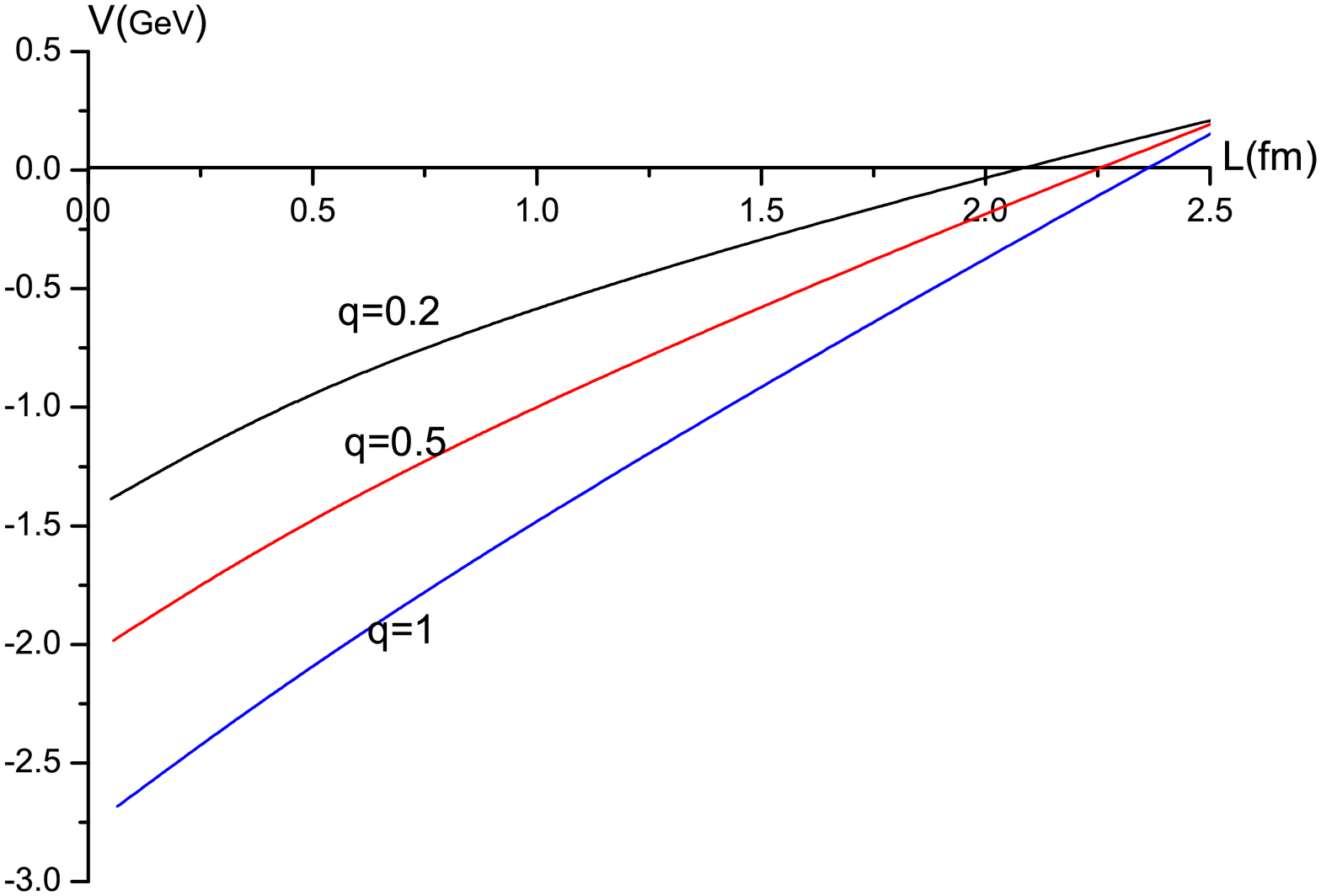}
\caption{Plots of $V$ versus $L$ with three different values of
$q$. Here $T=0.1GeV$. From top to bottom $q=0.2,0.5,1.0$.}
\end{figure}

Before going further, we would like to discuss the values of some
parameters. The coefficient $\frac{1}{\pi\alpha'}$ and the AdS
radius $R$ do not play any role in the physical discussion, so we
set $\frac{1}{\pi\alpha'}=2R^2=1$, the similar assumption can be
found in \cite{YS1}. In addition, the value of $q$ is related to
the gluon condensation $<TrF^2>$ in boundary theory and non-zero
$q$ implies that the chiral symmetry is broken \cite{BG}, but in
the present work, we only consider the instanton density $q$ as an
external parameter, so we can choose the values of $q$ properly,
see also in \cite{MD}.

To find how the instanton density $q$ affects the heavy quark
potential qualitatively. In Fig 1 we plot the potential $V$ as a
function of the inter-quark distance $L$ with a fixed temperature
$T=0.1GeV$ for three different values of $q$. In the plots from
top to bottom $q=0.2,0.5,1.0$, respectively. From the figures, we
can see clearly that the potential decreases as $q$ increases.
Also, one finds that by increasing $q$ the dissociation length
increases. Therefore, the instanton density tends to suppress the
heavy quark potential as well as make the dissociation length
longer. Interestingly, some other corrections such as the
sub-leading order corrections \cite{ZQ} and the higher curvature
corrections \cite{KB1} both make the dissociation length shorter.

\section{jet quenching parameter}

Next, we investigate the jet quenching parameter in this
D-instanton background. The eikonal approximation relates the jet
quenching parameter with the expectation value of an adjoint
Wilson loop $W^A[{\cal C}]$, where ${\cal C}$ is a rectangular
contour of size $L\times L_-$, the sides with length $L_-$ run
along the light-cone, the limit $L_-\to\infty$ is taken in the
end. Under the dipole approximation, which is valid for small
transverse separation $L$, the jet quenching parameter defined in
Ref \cite{RB} can be extracted from the asymptotic expression for
$TL<<1$
\begin{equation}
<W^A[{\cal C}]> \approx \exp [-\frac{1}{4\sqrt{2}}\hat{q}L_-L^2],
\label{jet}
\end{equation}
where $<W^A[{\cal C}]>\approx <W^F[{\cal C}]>^2$ with $<W^F[{\cal
C}]>$ the thermal expectation value in the fundamental
representation.

Using the AdS/CFT correspondence, one can calculate $<W^F[{\cal
C}]>$ according to:
\begin{equation}
<W^F[{\cal C}]>\approx\exp[-S_I] \label{WF},
\end{equation}
with $S_I=S-S_0$, where $S$ is the total energy of the quark pair,
$S_0$ is the self-energy of the isolated quark and the isolated
anti-quark.

Thus, the general relation for the jet quenching parameter can be
written as
\begin{equation}
\hat{q}=8\sqrt{2}\frac{S_I}{L_-L^2}.\label{q}
\end{equation}

By virtue of the light-cone coordinate
$x^\mu=(r,x^+,x^-,x_2,x_3)$, the metric Eq.(\ref{metric}) becomes
\begin{equation}
ds^2=-e^{\frac{\Phi}{2}}\frac{r^2}{R^2}(1+f)dx^+dx^-+e^{\frac{\Phi}{2}}\frac{r^2}{R^2}(dx_2^2+dx_3^2)+e^{\frac{\Phi}{2}}\frac{1}{2}\frac{r^2}{R^2}(1-f)[(dx^+)^2+(dx^-)^2]+e^{\frac{\Phi}{2}}\frac{R^2}{r^2f}dr^2.
\label{metric1}
\end{equation}

As the Wilson loop in question stretches across $x_2$ and lies at
$x^+=constant,x_3=constant$, we can choose the static gauge as
\begin{equation}
x^-=\tau, \qquad x_2=\sigma,
\end{equation}
and suppose a profile of $r=r(\sigma)$, then Eq.(\ref{metric1})
becomes
\begin{equation}
ds^2=e^{\frac{\Phi}{2}}[\frac{1}{2}(\frac{r^2}{R^2}-f_1)d\tau^2+(\frac{r^2}{R^2}+\frac{\dot{r}^2}{f_1})d\sigma^2],
\end{equation}
where $\dot{r}=\frac{dr}{d\sigma}$,
$f_1\equiv\frac{r^2}{R^2}(1-\frac{r_t^4}{r^4})$.

Then the Nambu-Goto action is given by
\begin{equation}
S=\frac{\sqrt{2}L_-}{2\pi\alpha^\prime}\int_0^{\frac{L}{2}}d\sigma\sqrt{e^{\Phi}(\frac{r^2}{R^2}-f_1)(\frac{r^2}{R^2}+\frac{\dot{r}^2}{f_1})},\label{NGG}
\end{equation}
where the boundary condition is $r(\pm\frac{L}{2})=\infty$.

Note that the integrand doe not depend explicitly on $\sigma$, so
we have a conserved quantity
\begin{equation}
\frac{\partial\mathcal L}{\partial\dot{r}}\dot{r}-\mathcal
L=\frac{-e^{\Phi}(\frac{r^2}{R^2}-f_1)\frac{r^2}{R^2}}{\sqrt{e^{\Phi}(\frac{r^2}{R^2}-f_1)(\frac{r^2}{R^2}+\frac{\dot{r}^2}{f_1})}}=C,
\end{equation}
which leads to
\begin{equation}
\dot{r}^2=\frac{f_1r^2}{R^2C^2}[\frac{e^\Phi
r^2(\frac{r^2}{R^2}-f_1)}{R^2}-C^2].\label{rr}
\end{equation}

The above Eq.(\ref{rr}) involves determining the zeros and the
region of positivity of the right-hand side. It was argued
\cite{liu} that the turning point occurs at $f_1=0$, implying
$\dot{r}=0$ at the horizon $r=r_t$.

As a matter of convenience, we set $B\equiv\frac{1}{C^2}$. For the
low energy limit ($C\rightarrow0$), one can integrate
Eq.(\ref{rr}) to leading order in $\frac{1}{B}$ and obtain the
following relation
\begin{equation}
L=2R^2\int_{r_t}^\infty dr
\sqrt{\frac{1}{(\frac{r^2}{R^2}-f_1)Bf_1r^4e^\Phi}}.\label{LL}
\end{equation}

On the other hand, plugging Eq.(\ref{rr}) into Eq.(\ref{NGG}), one
can rewrite the Nambu-Goto action as follows:

\begin{eqnarray}
S&=&\frac{\sqrt{2}L_-}{2\pi\alpha^\prime}\int_{r_t}^\infty dr
\sqrt{\frac{e^{2\Phi}(\frac{r^2}{R^2}-f_1)^2r^2}{f_1[e^\Phi
r^2(\frac{r^2}{R^2}-f_1)-R^2C^2]}}\nonumber\\
&=&\frac{\sqrt{2}L_-\sqrt{B}}{2\pi\alpha^\prime}\int_{r_t}^\infty
dr\frac{e^\Phi(\frac{r^2}{R^2}-f_1)r}{\sqrt{e^\Phi(\frac{r^2}{R^2}-f_1)Bf_1r^2-f_1R^2}}.\label{NGG1}
\end{eqnarray}

Likewise, we expand Eq.(\ref{NGG1}) to leading order of
$\frac{1}{B}$,
\begin{equation}
S=\frac{\sqrt{2}L_-}{2\pi\alpha^\prime}\int_{r_t}^\infty
dr[1+\frac{R^2}{2e^\Phi(\frac{r^2}{R^2}-f_1)Br^2}]
\sqrt{\frac{1}{f_1}e^\Phi(\frac{r^2}{R^2}-f_1)} .
\end{equation}

This action needs to be subtracted by the self energy of the two
quarks, that is
\begin{eqnarray}
S_0&=&\frac{2L_-}{2\pi\alpha^\prime}\int_{r_t}^\infty
dr\sqrt{g_{--}g_{rr}}\nonumber\\
&=&\frac{\sqrt{2}L_-}{2\pi\alpha^\prime}\int_{r_t}^\infty dr
\sqrt{\frac{1}{f_1}e^\Phi(\frac{r^2}{R^2}-f_1)}.
\end{eqnarray}

The subtracted action is therefore:
\begin{equation}
S_I=S-S_0=\frac{\sqrt{2}L_-R^2}{4\pi\alpha^\prime
B}\int_{r_t}^\infty
dr\sqrt{\frac{1}{(\frac{r^2}{R^2}-f_1)f_1r^4e^\Phi}}.\label{SI}
\end{equation}

Thus, from Eq.(\ref{q}), Eq.(\ref{LL}) and Eq.(\ref{SI}), we can
obtain the jet quenching parameter in a D-instanton background
\begin{equation}
\hat{q}=\frac{J(q)^{-1}}{\pi\alpha^\prime},\label{q1}
\end{equation}
where
\begin{equation}
J(q)=R^2\int_{r_t}^\infty
dr\sqrt{\frac{1}{(\frac{r^2}{R^2}-f_1)f_1r^4e^\Phi}}.
\end{equation}

Note that one can recover the jet quenching parameter of $N=4$ SYM
theory \cite{liu} by plugging the instanton density $q=0$ in
Eq.(\ref{q1}).

Numerically, we plot the curve of $\hat{q}/\hat{q}_{SYM}$ in terms
of the instanton density $q$ at a fixed temperature in Fig 2. The
plot shows that the jet quenching parameter in a D-instanton
background is larger than that of $N=4$ SYM theory. Also we find
that the jet quenching parameter increases as the instanton
density increases. This result is in agreement with that in
\cite{JF} which argues that for $\mathcal N=4$ SYM theory certain
marginal deformations have the effect of enhancing $\hat{q}$.

Furthermore, we would like to compare the results with
experimental data. Before going on, we should discuss the
values for $\alpha^\prime$ and $\lambda$ at hand. We here take
$\alpha^\prime=0.5$, which is reasonable for temperatures not far
above the QCD phase transition \cite{liu}. In addition, the
typical interval of $\lambda$ is $5.5<\lambda<6\pi$ \cite{GB1}. We
now use $\alpha^\prime=0.5$, $\lambda=6\pi$ and $T=300Mev$ to make
estimates. From Eq.(\ref{q1}), we find
$\hat{q}=5.0,5.31,5.58GeV^2/fm$ for $q=1,2,3$. These values of the
jet quenching parameter are consistent with the extracted values
from RHIC data($5\to25GeV^2/fm$) \cite{JD}.

\begin{figure}
\centering
\includegraphics[width=8cm]{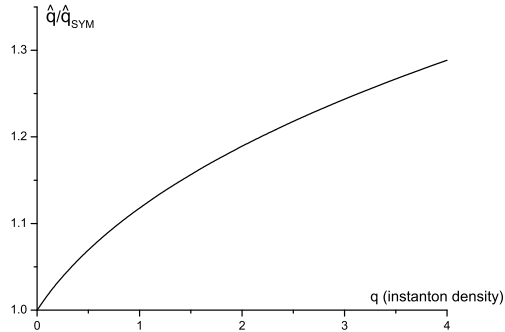}
\caption{Plots of $\hat{q}/\hat{q}_{SYM}$ versus instanton density
q at a fixed temperature. Here $T=300Mev$}
\end{figure}

\section{conclusion and discussion}

In this paper, we have investigated the heavy quark potential and
the jet quenching parameter in a D-instanton background. The dual
gravitational theory is related to a near horizon limit of stack
of black D3-branes with homogeneously distributed D-instantons.
Although the theory is not directly applicable to QCD, the
features of D3-D(-1) configuration is similar to QCD. Thus, one
can expect the results obtained from this theory should shed
qualitative insights into analogous questions in QCD.

In section III, we have investigated the heavy quark potential in
this D-instanton background. The potential was obtained by
calculating the Nambu-Goto action of string attaching the
rectangular Wilson loop. It is shown that the presence of
instantons tends to suppress the heavy quark potential and
increase the dissociation length.

In section IV, the jet quenching parameter has been studied in
this D-instanton background as well. It is found that the nonzero
instanton density has the effect of enhancing the jet quenching
parameter. Also, after taking some proper values of
$\alpha^\prime$ and $\lambda$, we observe that the results are
consistent with the experimental data.

Finally, it is interesting to note that the instanton effects on
the heavy-quark potential has also been studied from lattice QCD
recently \cite{BT}.

\section{Acknowledgments}

This research is partly supported by the Ministry of Science and
Technology of China (MSTC) under the ¡°973¡± Project no.
2015CB856904(4). Zi-qiang Zhang is supported by the NSFC under
Grant no. 11547204. Gang Chen is supported by the NSFC under Grant
no. 11475149. De-fu Hou is partly supported by the NSFC under Grant
nos. 11375070, 11221504 and 11135011.

%%%%%%%%%%%%%%%%%%%%%%%%%%%%%%%%%%%%%%%%

\end{document}